\documentclass[american,english,british,aps,prl,superscriptaddress,showpacs,twocolumn,nofootinbib]{revtex4-1}
\usepackage[latin9]{inputenc}
\usepackage{color}
\usepackage{babel}
\usepackage{amstext}
\usepackage{amssymb}
\usepackage{graphicx}
\usepackage[unicode=true,pdfusetitle,
 bookmarks=true,bookmarksnumbered=false,bookmarksopen=false,
 breaklinks=False,pdfborder={0 0 0},pdfborderstyle={},backref=false,colorlinks=False]
 {hyperref}
\makeatletter
\renewcommand\[{\begin{equation}}
\renewcommand\]{\end{equation}} 

\renewcommand{\fnum@figure}{FIG.~\thefigure}
\newcommand\blfootnote[1]{%
  \begingroup
  \renewcommand\thefootnote{}\footnote{#1}%
  \addtocounter{footnote}{-1}%
  \endgroup
}

\makeatother

\begin{document}

\title{Site-selective measurement of coupled spin pairs in an organic semiconductor}

\author{S. L. Bayliss$^{\dagger,*}$}
\selectlanguage{american}%

\address{LPS, Univ. Paris-Sud, CNRS, UMR 8502, F-91405, Orsay, France}
\selectlanguage{english}%

\address{Cavendish Laboratory, J. J. Thomson Avenue, University of Cambridge,
Cambridge CB3 0HE, UK }

\address{Berlin Joint EPR Lab, Fachbereich Physik, Freie Universität Berlin,
D-14195 Berlin, Germany}
\selectlanguage{british}%

\author{L. R. Weiss$^{\dagger}$}
\selectlanguage{english}%

\address{Cavendish Laboratory, J. J. Thomson Avenue, University of Cambridge,
Cambridge CB3 0HE, UK }
\selectlanguage{british}%

\author{A. Mitioglu}
\selectlanguage{english}%

\address{High Field Magnet Laboratory (HFML-EMFL), Radboud University, 6525
ED Nijmegen, The Netherlands}
\selectlanguage{british}%

\author{K. Galkowski}
\selectlanguage{english}%

\address{Laboratoire National des Champs Magnetiques Intenses, CNRS-UJF-UPS-INSA,
143 Avenue de Rangueil, 31400 Toulouse, France}
\selectlanguage{british}%

\author{Z. Yang}
\selectlanguage{english}%

\address{Laboratoire National des Champs Magnetiques Intenses, CNRS-UJF-UPS-INSA,
143 Avenue de Rangueil, 31400 Toulouse, France}
\selectlanguage{british}%

\author{K. Yunusova}
\selectlanguage{american}%

\address{LPS, Univ. Paris-Sud, CNRS, UMR 8502, F-91405, Orsay, France}
\selectlanguage{british}%

\author{A. Surrente}
\selectlanguage{english}%

\address{Laboratoire National des Champs Magnetiques Intenses, CNRS-UJF-UPS-INSA,
143 Avenue de Rangueil, 31400 Toulouse, France}
\selectlanguage{british}%

\author{K. J. Thorley}
\selectlanguage{english}%

\address{Department of Chemistry, University of Kentucky, Lexington, KY 40506-0055,
USA }
\selectlanguage{british}%

\author{J. Behrends}
\selectlanguage{english}%

\address{Berlin Joint EPR Lab, Fachbereich Physik, Freie Universität Berlin,
D-14195 Berlin, Germany}
\selectlanguage{british}%

\author{R. Bittl}
\selectlanguage{english}%

\address{Berlin Joint EPR Lab, Fachbereich Physik, Freie Universität Berlin,
D-14195 Berlin, Germany}
\selectlanguage{british}%

\author{J. E. Anthony}
\selectlanguage{english}%

\address{Department of Chemistry, University of Kentucky, Lexington, KY 40506-0055,
USA }
\selectlanguage{british}%

\author{A. Rao}
\selectlanguage{english}%

\address{Cavendish Laboratory, J. J. Thomson Avenue, University of Cambridge,
Cambridge CB3 0HE, UK }
\selectlanguage{british}%

\author{R. H. Friend}
\selectlanguage{english}%

\address{Cavendish Laboratory, J. J. Thomson Avenue, University of Cambridge,
Cambridge CB3 0HE, UK }
\selectlanguage{british}%

\author{P. Plochocka}
\selectlanguage{english}%

\address{Laboratoire National des Champs Magnetiques Intenses, CNRS-UJF-UPS-INSA,
143 Avenue de Rangueil, 31400 Toulouse, France}
\selectlanguage{british}%

\author{P. C. M. Christianen}
\selectlanguage{english}%

\address{High Field Magnet Laboratory (HFML-EMFL), Radboud University, 6525
ED Nijmegen, The Netherlands}
\selectlanguage{british}%

\author{N. C. Greenham\foreignlanguage{english}{$^{\star}$}}
\selectlanguage{english}%

\address{Cavendish Laboratory, J. J. Thomson Avenue, University of Cambridge,
Cambridge CB3 0HE, UK }

\author{A. D. Chepelianskii$^{\star}$}
\selectlanguage{american}%

\address{LPS, Univ. Paris-Sud, CNRS, UMR 8502, F-91405, Orsay, France}
\selectlanguage{english}%
\begin{abstract}
From organic electronics to biological systems, understanding the
role of intermolecular interactions between spin pairs is a key challenge.
Here we show how such pairs can be selectively addressed with combined
spin and optical sensitivity. We demonstrate this for bound pairs
of spin-triplet excitations formed by singlet fission, with direct
applicability across a wide range of synthetic and biological systems.
We show that the site-sensitivity of exchange coupling allows distinct
triplet pairs to be resonantly addressed at different magnetic fields,
tuning them between optically bright singlet ($S=0$) and dark triplet,
quintet ($S=1,2$) configurations: this induces narrow holes in a
broad optical emission spectrum, uncovering exchange-specific luminescence.
Using fields up to 60~T, we identify three distinct triplet-pair
sites, with exchange couplings varying over an order of magnitude
(0.3-5~meV), each with its own luminescence spectrum, coexisting
in a single material. Our results reveal how site-selectivity can
be achieved for organic spin pairs in a broad range of systems.\foreignlanguage{british}{}
\end{abstract}
\maketitle
\selectlanguage{british}%
\blfootnote{$^\dagger$These authors contributed equally to this work.\\$^\star$alexei.chepelianskii@u-psud.fr, ncg11@cam.ac.uk. \\$^*$Present address: Institute for Molecular Engineering, University of Chicago, Chicago, Illinois 60637, USA} 

\selectlanguage{english}%
Spin pairs control the behavior of systems ranging from quantum circuits
to photosynthetic reaction centers \cite{petta2005coherent,lubitz2002radicals}.
In molecular materials, such pairs mediate a diverse range of processes
such as light emission, charge separation and energy harvesting \cite{Clarke2010,Steiner1989,mccamey2008spin}.
The relevant spin-pair may consist of two spin-1/2 particles, either
in the form of a bound exciton or weakly coupled electron-hole pair,
or spin-1 pairs, which have recently emerged as alternatives for efficient
light emission and harvesting \cite{Congreve2013,mezyk2009,liu2009,van2015kinetic}.
A key challenge in understanding and using such pairs is accessing
the local molecular environments which support their generation and
evolution within more complex structures, information which could
ultimately lead to active control of their properties.

\begin{figure*}
\begin{centering}
\textcolor{black}{\includegraphics[width=0.95\textwidth]{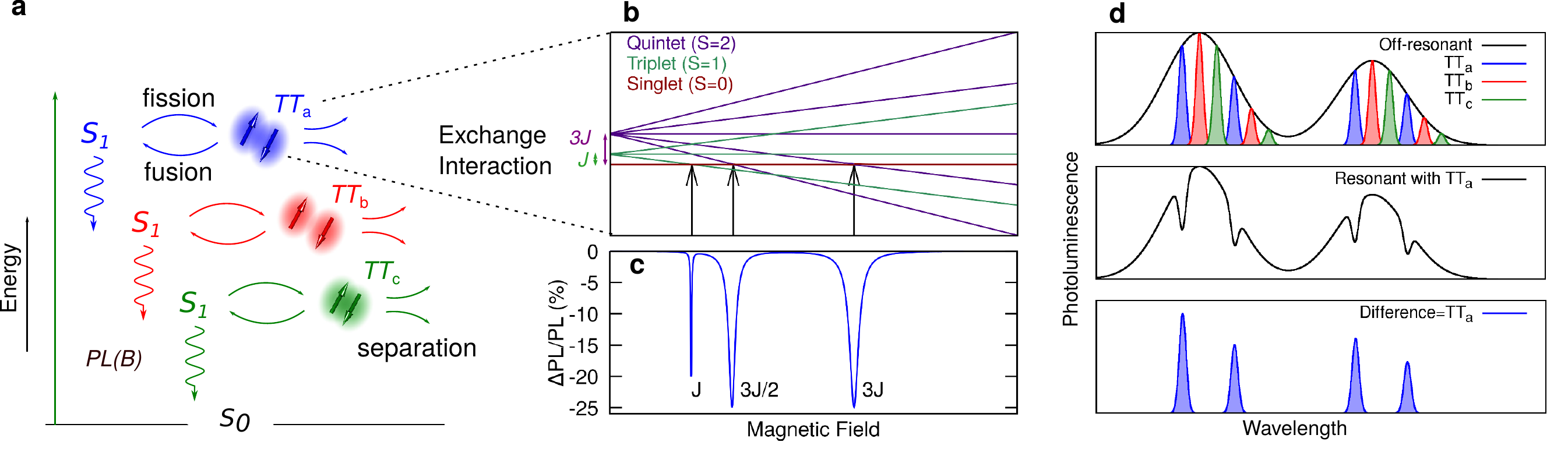}}
\par\end{centering}
\textcolor{black}{\caption{\label{fig:fig1}Selective addressing of exchange-coupled triplet-exciton
pairs. (\emph{A})\textbf{ }Schematic of spin-pair generation by singlet
fission for an ensemble of pair sites with different exchange interactions.
Photon absorption generates a spin-singlet exciton ($S_{1}$), which
can radiatively decay, producing photoluminescence (PL), or undergo
fission into a pair of triplet excitons $(TT)$. Fusion of this triplet
pair reforms the singlet exciton, while dissociation destroys it.
(\emph{B})/(\emph{C})\textbf{ }Triplet-pair level anticrossings for
a single exchange energy. A magnetic field tunes optically dark triplet
or quintet spin sub-levels into near-degeneracy with the bright singlet
state, resulting in selective reductions in the PL at fields proportional
to the exchange interaction $J$. $\Delta\mbox{PL/PL=[PL(B)-PL(0)]/PL(0)}$.
(\emph{D})\textbf{ }The magnetic field induced anticrossings create
spectral holes linked to specific triplet pair. This enables the narrow
associated emission profiles of triplet pairs with different exchange
interactions to be extracted.}
}
\end{figure*}

Here we demonstrate that the joint dependence of spin and electronic
interactions on pair conformation provides a handle to separate such
states and extract their discrete environments from a broader energetic
landscape. We apply this technique to measure distinct triplet-pairs
formed by singlet fission (Fig.~\ref{fig:fig1}\emph{A}), a process
which generates two spin $S=1$ excitons from a photogenerated $S=0$
singlet exciton, and is of great current interest for solar energy
conversion \cite{Smith2010,Thompson2014,Tabachnyk2014}. We simultaneously
extract the exchange energies and optical spectra of three different
triplet-pair sites within the same material. Using a magnetic field,
we tune different triplet pairs into excited-state avoided crossings,
which we detect as spectral holes in an inhomogeneously broadened
photoluminescence (PL) spectrum. This enables combined spin and optical
characterization of these states: the fields required to induce avoided
crossings directly measure the set of pair exchange-coupling strengths,
while the spectral holes provide narrow, spin-specific optical profiles
of the states. We extract multiple triplet-pair states with exchange
couplings varying by an order of magnitude and decouple their distinct
luminescence spectra from an otherwise inhomogeneously broadened background,
reaching sub-nm spectral linewidths. Our results open up new means
of determining structure-function relations of coupled spins and identify
unambiguous pair signatures. This approach is directly applicable
to a range of organic systems: from electron-hole pairs in next-generation
light-emitting diodes to coupled excitons in  light harvesters.\textcolor{black}{}

\subsection*{Selectively addressing exchange-coupled triplets}

Despite their key role in light-emitters and harvesters, triplet pairs
have only recently been discovered to form exchange-coupled states
\cite{weiss2016strongly,tayebjee2016quintet,basel2017unified,wakasa2015can,yago2016magnetic}.
We start by outlining how such states can be selectively addressed
to provide a site-specific measurement of their exchange interactions
and associated optical spectra. Here we describe the specific case
of singlet fission, but emphasise that the approach can be directly
translated to many other molecular systems since spin conserving transitions
are a general feature of such materials.

Fig.~\ref{fig:fig1}\emph{A} outlines the process of triplet-pair
generation by singlet fission, where both fission and the subsequent
fusion process are spin-conserving. This makes the spectral regions
associated with triplet pairs sensitive to their spin states, which
can be resonantly tuned with an external magnetic field (Fig.~1\emph{B}).
For strongly exchange-coupled triplets, the eigenstates at zero magnetic
field consist of the pure singlet ($S=0$), triplet ($S$=1) and quintet
($S=2$) pairings of the two particles. Due to its singlet precursor,
fission selectively populates the $S=0$ triplet-pair configuration,
which is energetically separated from the optically inactive triplet
or quintet states due to the exchange interaction. Application of
a magnetic field enables these triplet or quintet states to be tuned
into resonance with the optically active singlet pair state when the
Zeeman energy matches the singlet-triplet or singlet-quintet exchange
splitting. At these field positions, bright singlet pair states become
hybridized with a dark triplet or quintet pair-state, manifesting
as a resonant reduction in the relevant PL spectral window (Fig.~\ref{fig:fig1}\emph{C})
\cite{wakasa2015can,bayliss2016spin,yago2016magnetic}. 

Crucially, the crossings directly address pairs with a specific exchange
coupling.For an exchange interaction $J\mathbf{S}_{1}\cdot\mathbf{S}_{2}$,
where $\mathbf{S}_{1,2}$ are the spin operators for the two triplets,
the resonances occur at $|J|$ (singlet-triplet crossing), and $3|J|/2$,
$3|J|$ (singlet-quintet crossings), giving a direct measurement of
the exchange. (Here  we take $J>0$ - see Supplementary Information.)
Furthermore, only the emission linked to the resonant triplet pair
will be diminished at each level crossing. The magnetic field resonances
will therefore selectively burn spectral holes linked to pairs with
a given exchange coupling (Fig.~\ref{fig:fig1}\emph{D}). From these
resonant spectral changes, both the spin and optical properties of
pair sites are therefore reconstructed. Importantly, since triplet
pairs with different exchange interactions will have separated resonant
fields, their associated spectra can be individually measured. Specific
spin-pairs with distinct spectral and spin properties can therefore
be disentangled in an ensemble measurement and their local environment
and microscopic properties probed. This is the key principle of our
approach to provide a spin- and site-selective measurement of coupled
organic spins.

\selectlanguage{british}%
\selectlanguage{english}%

\subsection*{TIPS-tetracene}

\begin{figure}
\begin{centering}
\textcolor{black}{\includegraphics[width=0.97\columnwidth]{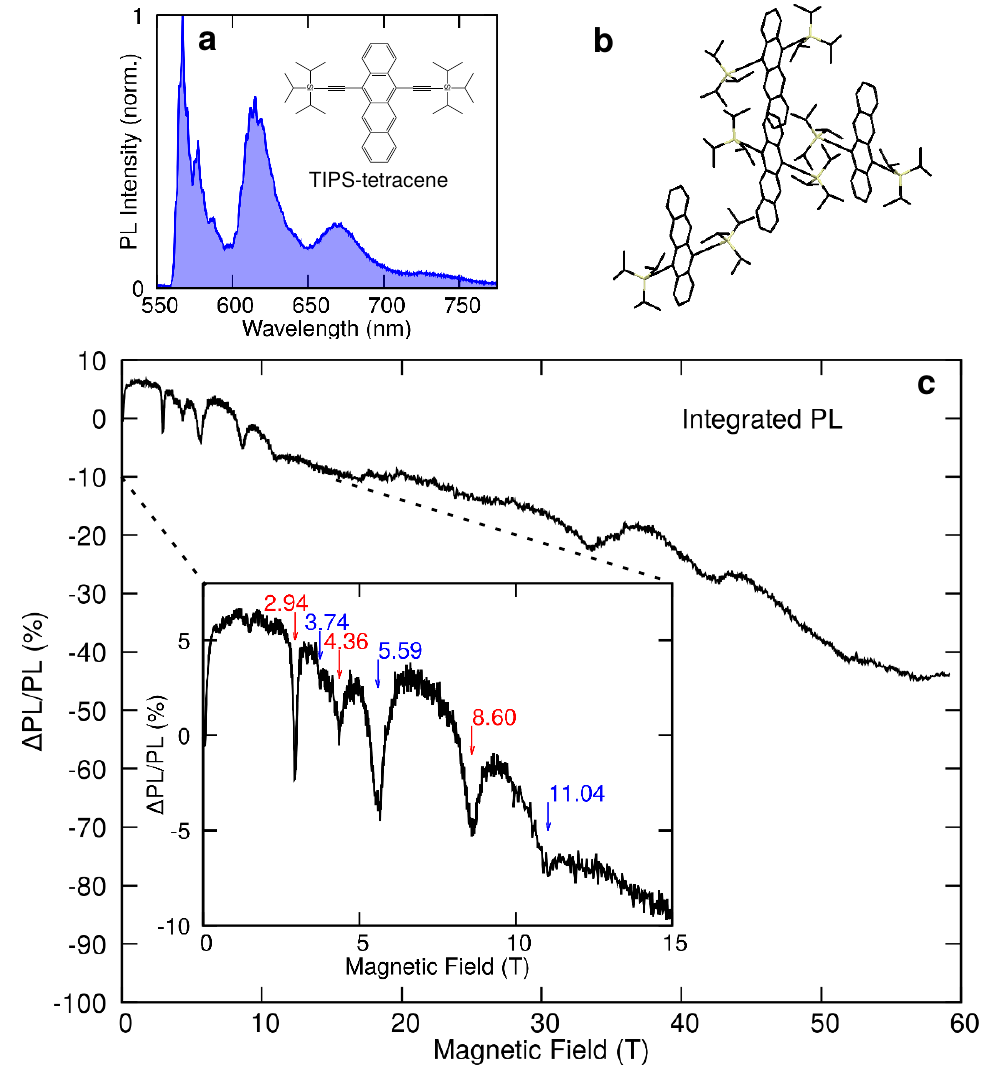}}
\par\end{centering}
\textcolor{black}{\caption{\label{fig:fig2}Level anticrossings of spin-1 pairs. (\emph{A})\textbf{
}Chemical schematic and photoluminescence spectrum of TIPS-tetracene
at 1.4 K (Sample 1). (\emph{B})\textbf{ }TIPS-tetracene unit cell
displaying four inequivalent molecules. (\emph{C})\textbf{ }Magneto-PL
at 1.4 K integrated across all wavelengths showing a series of resonances
(Sample 1, pulsed fields).}
}
\end{figure}

Of the expanding class of singlet fission materials for photovoltaic
applications, solution-processable systems with a triplet energy close
to the bandgap of silicon are particularly important since they could
be integrated directly with established high-efficiency silicon technologies.
One such material is TIPS-tetracene (Fig.~\ref{fig:fig2}\emph{A}/\emph{B}),
a solution-processable derivative of the archetypal fission material
tetracene \cite{Merrifield,Burdett2013a}, which has been shown to
undergo effective fission and generate exchange-coupled triplet pairs
\cite{stern2015identification,stern2017ultrafast,weiss2016strongly}.
Furthermore, singlet and triplet-pair states are nearly iso-energetic
in TIPS-tetracene, and so photoluminescence can be used to interrogate
the fission products \cite{Bayliss2014b,stern2015identification,stern2017ultrafast}.
Here we use TIPS-tetracene to study the spin and electronic structure
of coupled triplet excitons. To achieve both high spectral and field
resolution, we perform measurements using both pulsed ($<60\,\mbox{T}$)
and cw ($<33\,\mbox{T})$ magnetic fields on three identically prepared
samples (Methods): Sample 1 under pulsed fields at 1.4~K, and samples
2 and 3 under cw fields at 2 and 1.1~K respectively. Samples are
crystallites of $\sim\text{mm}$ linear dimensions, containing multiple
domains, prepared by evaporation from saturated solution. Samples
were not specifically oriented with respect to the magnetic field.
We first identify triplet-pair level anticrossings in TIPS-tetracene
and then use the avoided crossings to spectrally characterise multiple
distinct triplet pairs.

\textcolor{black}{}
\textcolor{black}{}
\begin{figure*}[t]
\begin{centering}
\textcolor{black}{\includegraphics[width=0.95\textwidth]{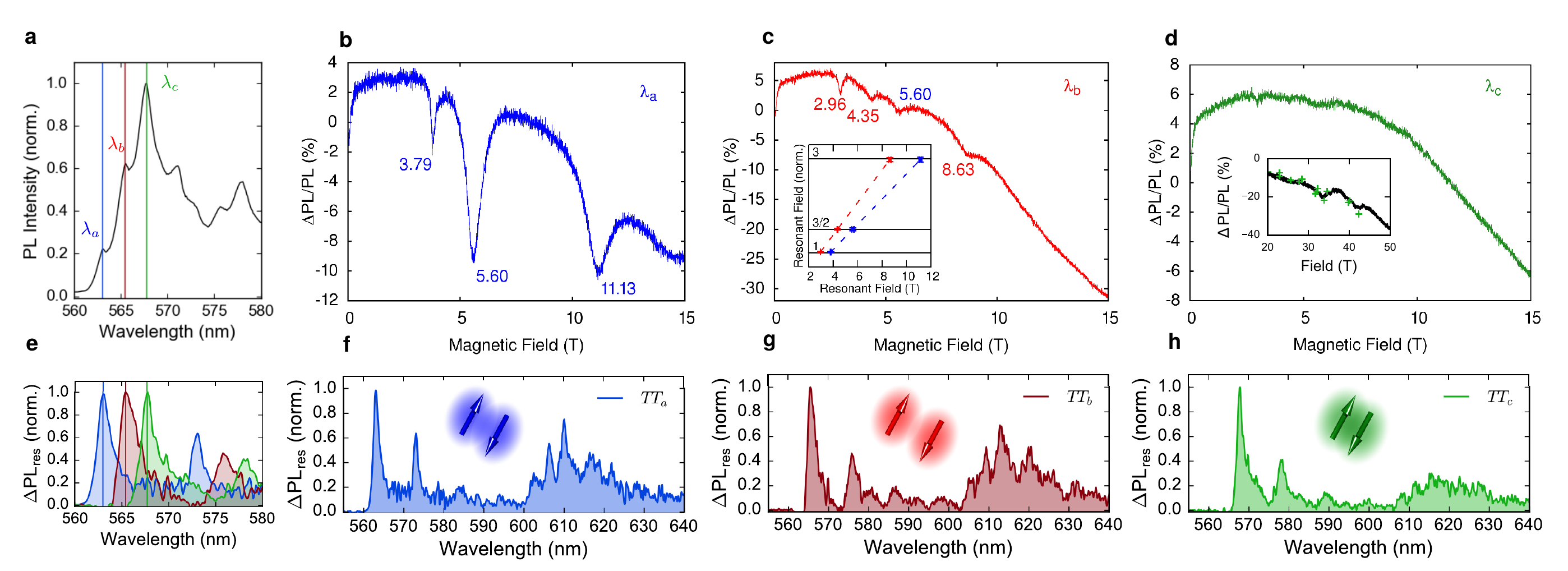}}
\par\end{centering}
\textcolor{black}{\caption{\label{fig:fig3}Magneto-optical spectroscopy of triplet pairs formed
by singlet fission. (\emph{A})\textbf{ }1.4~K PL spectrum with features
at $\lambda_{\mathrm{a}}=562.5\,\mbox{nm}$, $\lambda_{\mathrm{b}}=564.9\,\mbox{nm}$,
$\lambda_{\mathrm{c}}=567.1\,\mbox{nm}$ highlighted (Sample 1). (\emph{B})\emph{-}(\emph{D)}\textbf{
}Magneto-PL traces for the three spectral positions $\lambda_{\mathrm{a-c}}$
(Sample 2). (\emph{B})/(\emph{C}) Magnetic field resonances at $\lambda_{\mathrm{a}}$
and $\lambda_{\mathrm{b}}$ corresponding to triplet pairs with exchange
interactions of $0.44$ and $0.34\,\mbox{meV}$ ($J/g\mu_{B}=3.79,2.96\,\mbox{T}$).
(\emph{C})-inset, resonant fields appear with ratios 1:3/2:3, as expected
for the possible level crossings with the singlet state. Error bars
taken as 10 \% of the resonant linewidths. Dashed lines are guides
to the eye. (\emph{D})-inset.\textbf{ }S\textcolor{black}{pectrally
resolved PL measurements (marked field points) and integrated PL for
reference (solid line) at $\lambda_{\mathrm{c}}$. (}\textcolor{black}{\emph{E}}\textcolor{black}{)-(}\textcolor{black}{\emph{H}}\textcolor{black}{).}\textbf{\textcolor{black}{{}
}}\textcolor{black}{Extracted spectra for the triplet pairs associated
with the resonances (Sample 1): overlaid (}\textcolor{black}{\emph{E}}\textcolor{black}{),
and shown individually (}\textcolor{black}{\emph{F}}\textcolor{black}{)-}\textcolor{black}{\emph{(H}}\textcolor{black}{).}}
}
\end{figure*}

\subsection*{Triplet-pair level crossings}

Fig.~\ref{fig:fig2}\emph{C} shows the changes in integrated PL up
to 60~T for a TIPS-tetracene crystallite at $1.4\,\mbox{K}$ (Sample
1, pulsed fields - see Methods), where $\Delta$PL/PL={[}PL(B)-PL(0){]}/PL(0).
Below $1\,\mbox{T}$, the conventional singlet fission magnetic-field
effect is observed, indicative of weakly coupled triplet pairs \cite{Merrifield},
while at $\gtrsim1\,\mbox{T}$ a very different behavior arises. On
top of the monotonic PL reduction with field, which we discuss later,
multiple PL resonances are apparent: a series below 15~T, and additional
resonances above 30~T, indicating triplet-pair level anticrossings.
As shown in Fig.~\ref{fig:fig1}\emph{C}, for a given triplet pair
there are three possible resonances with the fission-generated singlet
state, occurring with field ratios 1:3/2:3. The number of resonances
in Fig.~\ref{fig:fig2}\emph{C} therefore indicates multiple triplet
pairs with different exchange interactions. While the resonances at
$<15\,\mbox{T}$ can be separated into two progressions with 1:3/2:3
field ratios (red/blue labels, Fig.~\ref{fig:fig2}\emph{C}) this
does not clearly assign them or associate them with particular optical
properties. We now show how the identified triplet-pair level crossings
can be used to unambiguously decouple and spectrally characterise
multiple interacting triplets in the same material.

\subsection*{Spectrally resolving interacting triplets}

Fig.~\ref{fig:fig3}\emph{B}-\emph{D} shows magneto-PL traces at
three different wavelengths $\lambda_{\mathrm{a,b,c}}$ which correspond
to high-energy regions of the TIPS-tetracene PL spectrum (Fig.~\ref{fig:fig3}\emph{A}).
In contrast to the integrated measurements, the magneto-PL at $\lambda_{\mathrm{a}}$
and $\lambda_{\mathrm{b}}$ shows a clear progression of three resonances
following the 1:3/2:3 field ratios expected for level anticrossings
with the singlet state (Fig.~\ref{fig:fig3}\emph{C},\emph{inset}),
giving exchange interactions of 0.44 and 0.34 meV respectively, i.e.
$J/g\mu_{B}=3.79,\,2.96\,\mbox{T}$ where $g\simeq2$ is the exciton
$g$-factor \cite{Bayliss2014b,weiss2016strongly} and $\mu_{B}$
the Bohr magneton. (Note that due to their spectral proximity, the
$5.6\,\mbox{T}$ $\lambda_{\mathrm{a}}$ resonance is also present
in the $\lambda_{\mathrm{b}}$ trace.) In contrast to the $\lambda_{\mathrm{a,b}}$
spectral positions, at $\lambda_{\mathrm{c}}$, resonances are present
only at much higher fields of 33.4 and $42.0\,\mbox{T}$. Since these
do not occur at the expected 1:3/2 field ratios, we assign them to
the lowest field - i.e. singlet-triplet - anticrossings of distinct
triplet pairs with exchange interactions of $3.87$ and $4.87\,\mbox{meV}$
respectively ($J/g\mu_{B}=33.4,\,42.0\,\mbox{T}$ ). This is further
supported by their distinct temperature dependences which we describe
later.

As outlined in Fig.~\ref{fig:fig1}\emph{D}, since the PL resonances
for triplet pairs with different exchange interactions are readily
separable in field, we can determine their emission characteristics
from the spectral components that are diminished at each resonant
field position i.e., the difference in PL ($\Delta\mbox{PL}_{\mathrm{res}}$)
when off-resonance vs. on resonance: $\Delta\mbox{PL}_{\mathrm{res}}=|\mbox{PL}(B_{\mathrm{res}.})-\mbox{PL}(B_{\mathrm{off\,res.}})|$.
(We note that for an accurate off-resonance subtraction in the presence
of more slowly changing non-resonant field effects, we take $\mbox{PL}(B_{\mathrm{off\,res.}})$
as the average of the spectra either side of the resonance.) The spectra
associated with each set of resonances (i.e. triplet pairs) are shown
in Fig.~\ref{fig:fig3}\emph{E}-\emph{H} and we label the associated
triplet pairs $TT_{\mathrm{a,b,c}}$. The resulting PL spectra show
similar vibronic progressions, yet shifted peak emission energies
with peaks centered at $\lambda_{\mathrm{a,b,c}}$ (Fig.~\ref{fig:fig3}\emph{E}).
The fact that the three spectra exhibit near-identical vibrational
progressions but with an overall shift relative to each other shows
that the states differ predominantly in their electronic rather than
vibrational coupling. The relative shift indicates a difference in
the local environment between the triplet pairs which results in distinct
electronic interactions with the surrounding molecules. The question
arises as to why a single material supports multiple triplet-pair
sites with distinguishable electronic and spin energy levels. A natural
explanation is the different molecular configurations accessible in
TIPS-tetracene in which there are four rotationally inequivalent molecules
in the crystal unit cell (Fig.~\ref{fig:fig2}\emph{B}) \cite{Bayliss2015b}.
Multiple triplet pairs may therefore be supported. Due to their differing
interaction strengths and electronic environments each pair is associated
with different exchange couplings and optical emission spectra. (We
note that as an alternative approach to species extraction, we find
that independent spectral decomposition algorithms show good agreement
with the spectra/lineshapes in Fig.~\ref{fig:fig3} - see Supplementary
Information.)\textcolor{black}{}
\begin{figure}[t]
\begin{centering}
\textcolor{black}{\includegraphics[width=0.95\columnwidth]{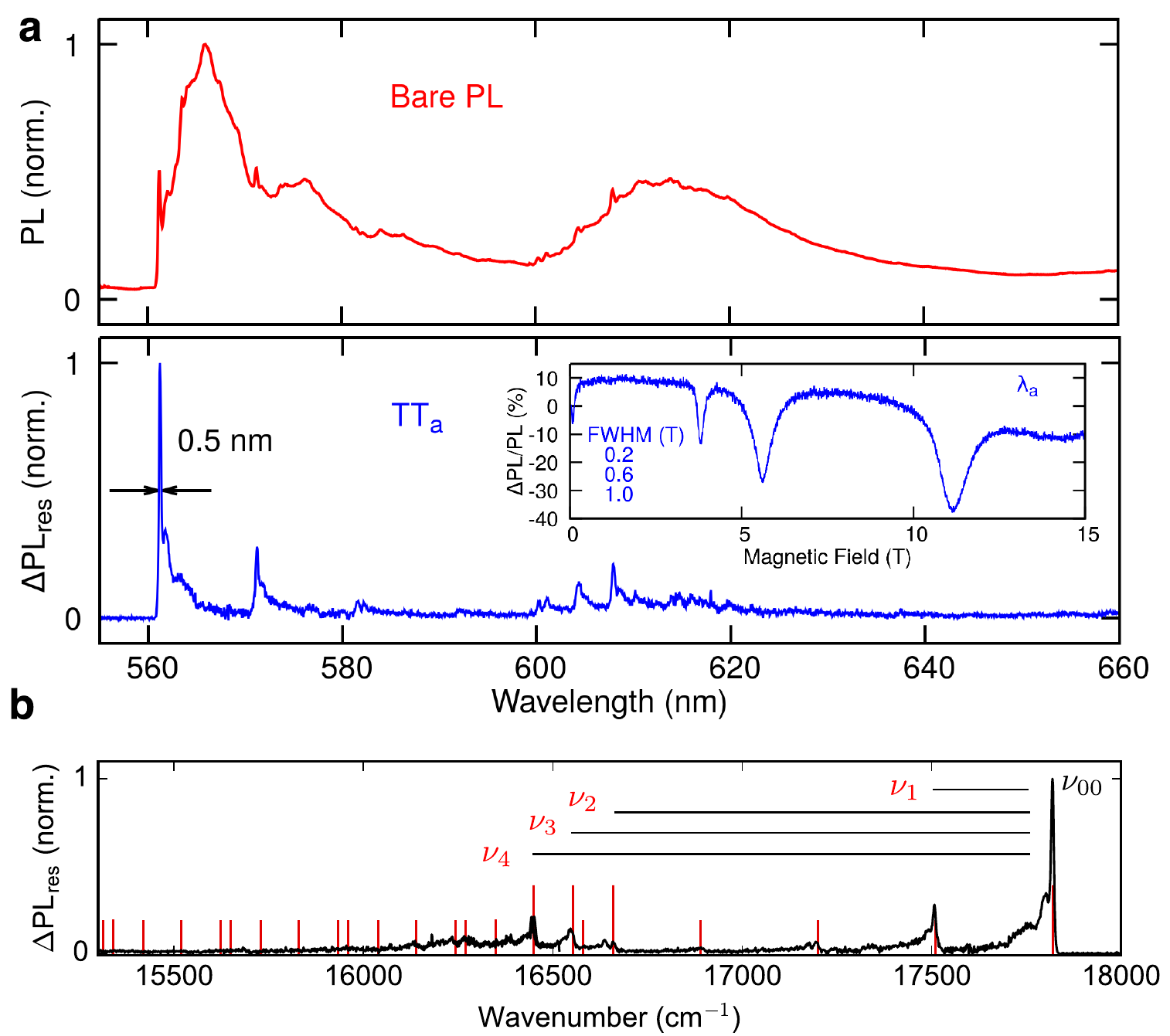}}
\par\end{centering}
\textcolor{black}{\caption{\label{fig:fig4}Vibrational structure in sub-nm resonant PL. (\emph{A})\textbf{
}Zero-field PL spectrum, $TT_{\mathrm{a}}$ spectrum extracted from
the 5.6~T resonance and PL resonances for Sample 3. (\emph{B})\textbf{
}Resonant PL spectrum of $TT_{\mathrm{a}}$ with idealised vibrational
progression (red lines) consisting of the 0-0 transition ($\nu_{00}$),
a dominant low-energy mode with wavenumber $\nu_{1}=310$~$\mbox{cm}^{-1}$
and three higher energy modes $\nu_{2,}\nu_{3},\nu_{4}=$$1160$,
$1270$, and $1370$~cm$^{-1}$.}
}
\end{figure}

\subsection*{Vibrational structure in $TT$ spectra}

In keeping with previous assignments, the first spectral peaks at
$\gtrsim560\,\mbox{nm}$ are attributed to 0-0, i.e. zero-phonon,
transitions \cite{stern2017ultrafast}. This is also consistent with
the greater overlap of low-energy modes on the higher-order vibrational
transitions described in detail below (Fig.~\ref{fig:fig4}\emph{B}).
Sample 3 (measured at the lowest temperature) exhibited pronounced
$TT_{\mathrm{a}}$ signatures (Fig.~\ref{fig:fig4}\emph{A}), with
linewidths of the extracted spectra reaching as low as 0.5~nm ($15\,\mbox{cm}^{-1}$),
significantly narrower than the $\sim10\,\mbox{nm}$ linewidth of
the 0-0 peak in the steady-state PL spectrum. This allows us to identify
the vibronic transitions shown in Fig.~\ref{fig:fig3} with greater
accuracy (Fig.~\ref{fig:fig4}\emph{B}). (Note that Sample 2 spectra
- Fig.~\ref{fig:fig3}\emph{F}-\emph{H} - were measured using a spectrometer
with lower spectral resolution, limiting the minimum linewidths).
We use this spectrum to extract four ground-state vibrational modes
involved in the emission process. Fig.~\ref{fig:fig4}\emph{B} shows
a stick spectrum of the progression of one lower energy mode with
wavenumber $\nu_{1}=310$ cm$^{-1}$, and three higher energy modes
($\nu_{2,}\nu_{3},\nu_{4}=$$1160$, $1270$, and $1370$~cm$^{-1}$),
showing good agreement with the measured spectra. These frequencies
are in agreement with modes found in the ground state Raman of TIPS-Tetracene
films \cite{stern2017ultrafast} with $\nu_{1}$ similar to typical
C-C-C out-of plane bending modes and $\nu_{2-4}$ similar to typical
C-C stretching/C-C-H bending modes \cite{alajtal2010effect}.

To our knowledge these are the first measurements of narrow optical
spectra which can be associated with triplet pairs. The sub-nm optical
linewidths obtained here are comparable to those obtained in fluorescence
line narrowing experiments of tetracene \cite{hofstraat1989temperature},
highlighting the sensitivity of this approach. In contrast to all-optical
measurements, the spin-sensitivity afforded here allows clear assignment
to triplet pairs. In addition, we note that spectral extraction of
triplet-pair signatures does not require clearly visible peaks in
the bare PL spectrum. For example, the longer wavelength peaks associated
with $TT_{\mathrm{a-c}}$ are unclear in the bare PL, and spectral
decomposition of $TT$ signatures is possible even in a sample with
barely visible $\lambda_{\mathrm{a},\mathrm{b}}$ peaks (Supplementary
Information).

\textcolor{black}{}
\begin{figure}[t]
\centering{}\textcolor{black}{\includegraphics[width=0.95\columnwidth]{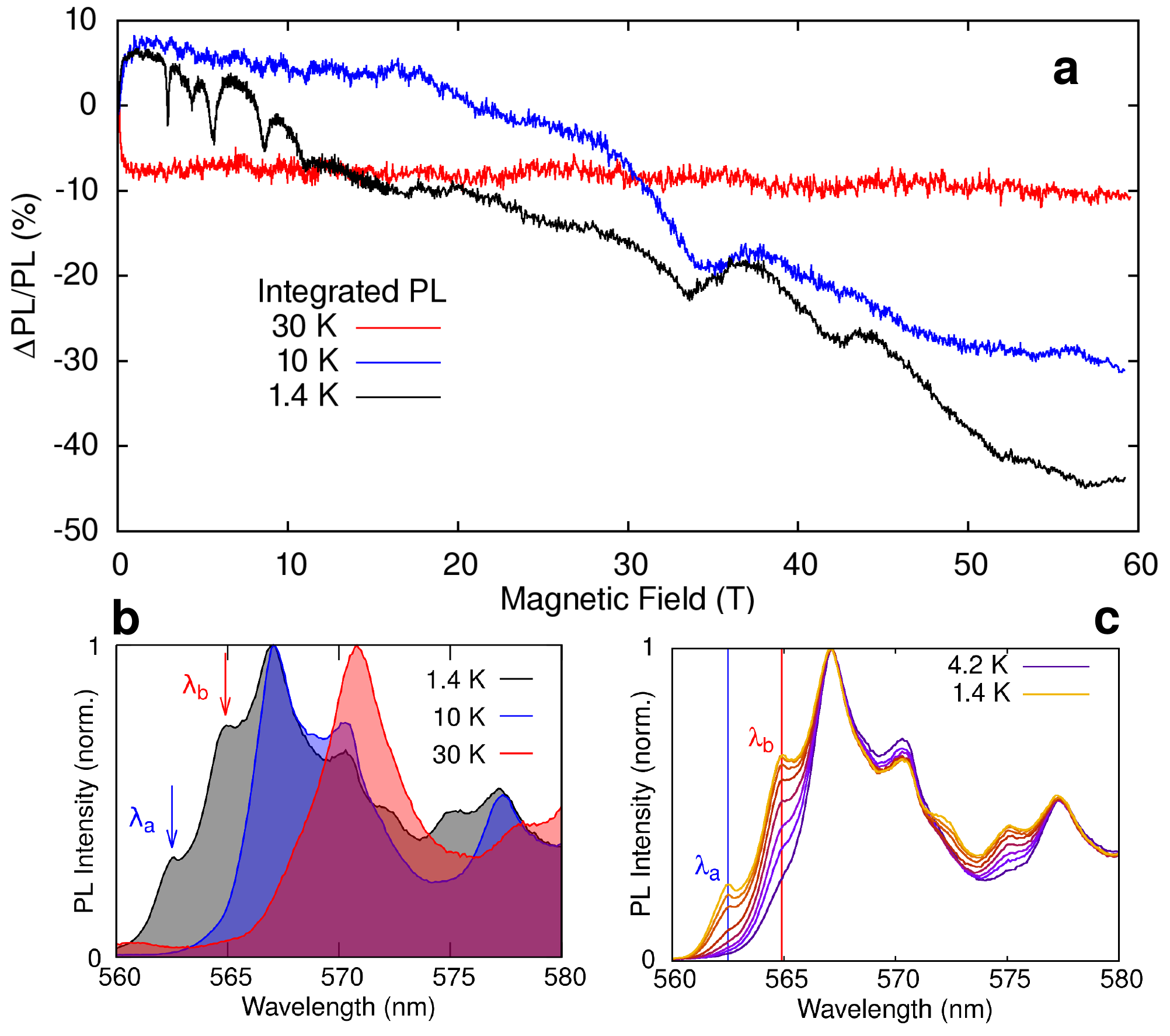}\caption{\label{fig:fig5}Temperature dependence of triplet-pair signatures.
(\emph{A})/(\emph{B})\textbf{ }Temperature-dependent integrated PL
traces and spectra (Sample 1).\textbf{ }(\emph{C})\textbf{ }Low-temperature
behavior of the $\lambda_{\mathrm{a}}$ and $\lambda_{\mathrm{b}}$
spectral features which correspond to the triplet pairs with $J/g\mu_{B}=3.79$
and 2.96 T respectively (Sample 1).}
}
\end{figure}

\subsection*{Temperature-dependent $TT$ signatures}

The identified triplet-pair species are further distinguishable through
their temperature dependences. Fig.~\ref{fig:fig5}\emph{A} shows
the temperature dependence of the resonances in the integrated PL
and the corresponding evolution of the emission spectrum. By 10 K,
the resonances below 15~T are lost, concurrent with the loss of the
$\lambda_{\mathrm{a}}$ and $\lambda_{\mathrm{b}}$ spectral features
(Fig.~\ref{fig:fig5}\emph{B}). The fact that the resonances at $\simeq$
33 and 42~T have distinct temperature dependences supports their
assignment to the first crossing of different triplet pairs (rather
than a single species with a more structured exchange interaction
\cite{kollmar1982theory}). By 30~K, no PL resonances are observed,
with no magnetic-field effect beyond $\sim1$~T. Measurement of PL
spectra between 4.2-1.4~K (Fig.~\ref{fig:fig5}\emph{C}) shows that
the $\lambda_{\mathrm{a},\mathrm{b}}$ spectral features evolve significantly
over this temperature range, indicating that escape from the associated
emission sites has an activation temperature on the order of a few
Kelvin ($\sim0.1\,\mbox{meV}$). Interestingly, this is approximately
the exchange coupling for $TT_{\mathrm{a},\mathrm{b}}$. However,
we note that this energy scale may alternatively be: (\emph{i}) a
reorganisation energy due to molecular reconfiguration or (\emph{ii})
an electronic barrier between different excited states\@.

\subsection*{High-field  spin mixing}

\begin{figure}[t]
\begin{centering}
\textcolor{black}{\includegraphics[width=1\columnwidth]{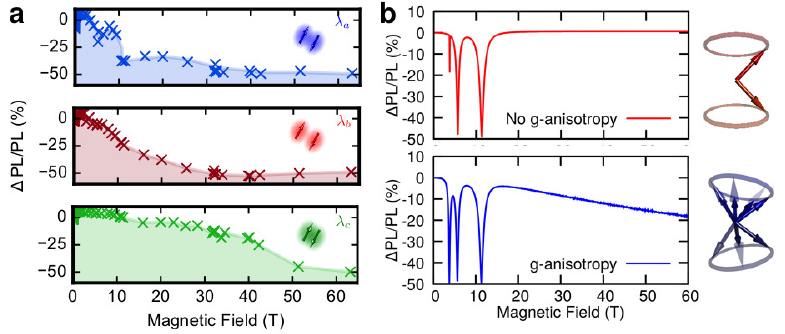}}
\par\end{centering}
\textcolor{black}{\caption{\label{fig:fig6}Triplet-pair spin-mixing. (\emph{A}) Spectrally resolved
high-field effect (Sample 1) showing $\Delta\mbox{PL/PL}$ at spectral
positions $\lambda_{\mathrm{a}-\mathrm{c}}$.\textbf{ }(\emph{B})
Simulation of the role of $g$-anisotropy. Inclusion of an anisotropic
$g$-factor enhances the singlet-triplet level crossing (at field
$J/g\mu_{B}=3.8\,\mbox{T}$) and produces a monotonic reduction in
PL with field.}
}
\end{figure}

While resonant spectral analysis provides a window into the electronic
structure associated with triplet pairs, the magnetic lineshapes provide
insight into spin-mixing mechanisms and the emissive species. The
magneto-PL shows a monotonic decrease with field, up to nearly 50~\%
at 60~T (Fig.~\ref{fig:fig2}\emph{C}), a drastically higher field
than the $<0.5\,\text{T}$ scale usually seen in organic systems.
This unanticipated high-field effect can be explained due to $g$-factor
anisotropy which can non-resonantly mix the singlet $|S\rangle$ and
$m=0$ triplet state $|T_{0}\rangle$, when triplets are orientationally
inequivalent, analogous to $\Delta g$ effects observed in spin-1/2
pairs due to differences in isotropic $g$-values \cite{Steiner1989,devir2014short,wang2016immense}.
The competition between spin-mixing $\Delta\mathbf{g}$ Hamiltonian
terms and total-spin-conserving exchange terms sets a characteristic
saturation field for the effect $\propto J/\Delta g_{\mathrm{eff}}$,
where $\Delta g_{\mathrm{eff}}$ is the relevant effective $g$-factor
difference (Supplementary Information). Triplet pairs with a larger
exchange interaction should therefore have a larger characteristic
field scale for this effect and hence also be distinguishable through
their non-resonant spin-mixing. Fig.~\ref{fig:fig6}\emph{C} shows
$\Delta$PL/PL for the three different spectral regions $\lambda_{\mathrm{a-c}}$
up to $68\,\mbox{T}$. The $\lambda_{\mathrm{a,b}}$ traces, which
correspond to triplet pairs with similar exchange interactions (0.44
and $0.34\,\mbox{meV}$) show a similar non-resonant lineshape which
saturates around 30~T, while the $\lambda_{\mathrm{c}}$ trace, associated
with an order of magnitude larger exchange interaction shows a much
higher characteristic field scale for PL reduction, consistent with
this mechanism. \foreignlanguage{british}{We note that high-field
effects have rarely been observed in organic materials in general,
and our observations show the relevance of spin-orbit coupling (responsible
for $g$-anisotropy), which is usually assumed to be negligible.}

\subsection*{Singlet-triplet level-crossings}

A difference in $g$-matrices also provides a mixing mechanism for
the singlet-triplet crossings. Since the pure $S=1$ triplet-pair
states are antisymmetric with respect to particle-exchange, while
the $S=0,2$ states are symmetric \cite{Merrifield}, different mixing
mechanisms are required for singlet-triplet vs. singlet-quintet hybridization.
Singlet-quintet mixing can be mediated by the intratriplet zero-field
splitting interaction \cite{bayliss2016spin} which characterises
the dipolar interaction between electron and hole and has strength
$D/g\mu_{B}=50\,\mbox{mT}$ in TIPS-tetracene \cite{JohnA.Weil,Bayliss2014b,Bayliss2015b,weiss2016strongly}.
However, to first order this coupling, leaves the singlet-triplet
crossing forbidden (Supplementary Information). Clear singlet-triplet
crossings seen for $TT_{\mathrm{a},\mathrm{b}}$ therefore indicate
an additional mixing mechanism. As with the high-field effect, this
can be provided by a $\Delta\mathbf{g}$ Hamiltonian term which mixes
singlet and triplets to first order (Fig.~\ref{fig:fig6}\emph{B}
and Supplementary Information) with strength $\sim\Delta g_{\mathrm{eff}}^{\prime}B\sim10^{-3}B$
for an expected $\Delta g_{\mathrm{eff}}^{\prime}\sim10^{-3}$ \cite{schott2017tuning}.
Additionally, this crossing can be mediated by hyperfine interactions,
with typical strengths of $\sim$mT in organic semiconductors \cite{mccamey2010hyperfine,zarea2014spin}. 

\subsection*{The role of kinetics in magnetic field effect}

Interestingly, the magnetic linewidths of the PL resonances (Fig.~\ref{fig:fig4}\emph{A})
are larger than expected based purely on the mixing matrix elements
for the crossings, which would give linewidths of $\lesssim50\,\mbox{mT}$.
We obtained similar linewidths in a single crystal sample (Supplementary
Information), and therefore a distribution in $J$ can be ruled out
as the dominant line broadening mechanism. Instead, as detailed in
the Supplementary Information, this indicates the resonance are predominantly
broadened by the kinetics of the fission/fusion process.
For both resonant and non-resonant PL reductions, mixing is predominantly
between the singlet, and one other (triplet or quintet) pair state,
and this sets a maximum $\Delta$PL/PL of $\simeq50\,\%$ (neglecting
annihilation to a single triplet). This maximum is based on the distribution
of $S=0$ character across one state at zero-field, vs. two states
at resonant positions/high field \cite{bayliss2016spin}. The fact
that the PL can be reduced by nearly 50 \% by a magnetic field (Fig.~\ref{fig:fig2}\emph{C},
Fig.~\ref{fig:fig4}\emph{A}) therefore indicates that strongly coupled
triplet pairs can dominate the steady-state emission properties of
singlet-fission systems. For identifying singlet fission, the observation
that exchange-coupled triplets can dominate steady-state magnetic
field effects is highly significant. Often, a low-field effect ($\lesssim100\,\mbox{mT}$)
characteristic of weakly coupled triplets \cite{Merrifield} is taken
to be a signature of the fission process \cite{Congreve2013}. In
contrast, our results show that singlet fission magnetic field effects
can be drastically different between strongly and weakly coupled triplets,
and that high-field effects ($\gtrsim1\,\mbox{T}$) can instead dominate.

We note that for fission generated triplet pairs the emissive species
may either be a distinct singlet exciton or, as proposed recently
\cite{stern2017ultrafast,yong2017entangled}, the triplet pairs themselves.
While typically challenging to distinguish these scenarios, the  combination
of kinetically broadened linewidths and near 50~\% resonant PL reductions
naturally arises only when triplet pairs emit via a separate singlet
state, rather than directly themselves, showing the additional utility
of these measurements in distinguishing these kinetic scenarios (Supplementary
Information).

\subsection*{Outlook}

The magneto-optic resolution of organic triplet pairs opens up the
possibility to correlate their exchange and electronic structure with
their chemical environment and physical conformation. Since the mixing
matrix elements relevant for the PL resonances depend on the relative
orientation between the external field and the triplet pair \cite{bayliss2016spin},
measuring orientationally dependent PL resonances should allow triplet
pairs to be assigned to specific molecular configurations.  Identification
of unambiguous spectral signatures of triplet pairs further means
that these states can now be studied through purely optical means.
For example, triplet-pair microscopy could be used to obtain information
on the spatial distribution of pair sites across microcrystalline
domains and map their diffusion \cite{irkhin2011direct,akselrod2014visualization,wan2015cooperative},
and resonant excitation could be used to address specific triplet
pairs through site-specific fluorescence \cite{bassler1999site,orrit1993high}\emph{.}

While here we spectrally resolve triplet pairs in a singlet fission
material, these results are applicable to a range of other organic
spin-pair systems. For example, triplet-triplet encounters are pivotal
in photovoltaic upconversion systems \cite{Singh-Rachford2010} and
organic light-emitting diodes \cite{baldo2000transient,van2015kinetic},
and triplet-pair level anticrossings should also be observable in
photovoltaic device architectures, where resonances could be measured
through solar-cell photocurrent or quantum-dot emission \cite{Thompson2014}.
 In spin-1/2 pairs, analogous spectrally resolved level crossings
should help to clarify the spin and electronic structure of the emissive
species central to thermally activated delayed fluorescence in next-generation
organic light-emitting diode materials, and extracting optical signatures
from level-crossings observed in synthetic and biological radical
pairs should provide further insights into these key intermediates
\cite{weiss2003direct,hiscock2016quantum,zarea2014spin,Steiner1989}.
Finally, the nanoscale sensitivity of exchange-coupled spins opens
up the possibility to deliberately engineer them as joint spin-optical
probes of complex molecular systems.

\medskip{}

\selectlanguage{british}%
\noindent \textbf{\emph{Methods}}  \\
\noindent Samples were excited by 532, 514 or 485 nm laser illumination (similar results
were obtained across this wavelength range). A long-pass filter was
used to remove the laser line, and the collected PL was either sent
to an avalanche photodiode for the integrated measurements or through
a monochromator to a nitrogen-cooled CCD for the spectrally resolved
measurements. Three different TIPS-tetracene crystallites prepared
by evaporation from saturated solution were used which we refer to
as samples 1-3. X-Ray diffraction confirmed that all samples indexed
to the same unit cell previously determined for TIPS-tetracene \cite{NIKLUQ},
demonstrating that they had the same underlying solid-state structure.
Integrated and spectrally resolved experiments to 68 T were performed
using Sample 1 under pulsed magnetic field at LNCMI Toulouse. Spectrally
resolved measurements up to 33 T were performed using samples 2 and
3 under steady-state fields at the HFML, Nijmegen. For low-temperature
measurements samples were either immersed in liquid helium (Samples
1 and 3) or cooled via exchange gas with a surrounding helium bath
(Sample 2), giving base temperatures of $\simeq1.4,\,2$ and $1.1$~K
for samples 1-3 respectively. PL spectra in Fig.~\ref{fig:fig5}\emph{C}
were taken with Sample 1 in helium under continuous pumping. Further
details and comparison of the samples are contained in the Supplementary
Information.

\selectlanguage{english}%
\medskip{}

\selectlanguage{english}%
\smallskip{}

\selectlanguage{british}%
\noindent \textbf{\emph{Acknowledgements}} 

\noindent This work was supported by HFML-RU/FOM and LNCMI-CNRS, members
of the European Magnetic Field Laboratory (EMFL) and by EPSRC (UK)
via its membership to the EMFL (grant no. EP/N01085X/1 and NS/A000060/1)
and through grant no. EP/M005143/1. L.R.W. acknowledges support of
the Gates-Cambridge and Winton Scholarships. We acknowledge support
from Labex ANR-10-LABX-0039-PALM, and ANR SPINEX. We are grateful
for support from DFG SPP-1601 (Bi-464/10-2).

\selectlanguage{english}%
\smallskip{}

\selectlanguage{british}%
\noindent \textbf{\emph{Author contributions}} 

\noindent S.L.B. and L.R.W. analysed the data and wrote the manuscript
with input from all authors. S.L.B., L.R.W., A.M. and K.Y. carried
out the experiments at the HFML. S.L.B., L.R.W., K.G., Z.Y., K.Y.,
A.S. and A.D.C. carried out the experiments at the LNCMI. K.J.T. and
J.E.A. provided the materials. All authors discussed the results.

\selectlanguage{english}%
\smallskip{}

\selectlanguage{british}%
\noindent \textbf{\emph{Additional information}} 

\noindent Correspondence should be addressed to A.D.C or N.C.G., and
requests for materials to J.E.A (anthony@uky.edu).

\selectlanguage{english}%
\smallskip{}

\selectlanguage{british}%
\noindent \textbf{\emph{Competing financial interests}} 

\noindent The authors declare no competing financial interests.\\
\vspace{10 mm}

\pagebreak
\widetext
\begin{center}
\textbf{\large Supplementary Information: Site-selective measurement of coupled spin pairs in an organic semiconductor}
\end{center}
\setcounter{equation}{0}
\setcounter{figure}{0}
\setcounter{table}{0}
\setcounter{page}{1}
\makeatletter
\renewcommand{\theequation}{S\arabic{equation}}
\renewcommand{\thefigure}{S\arabic{figure}}
\renewcommand{\thetable}{S\arabic{table}}
\renewcommand{\bibnumfmt}[1]{[S#1]}
\renewcommand{\citenumfont}[1]{S#1}

\begin{figure}[h!]
\centering{}\includegraphics[width=0.3\textwidth]{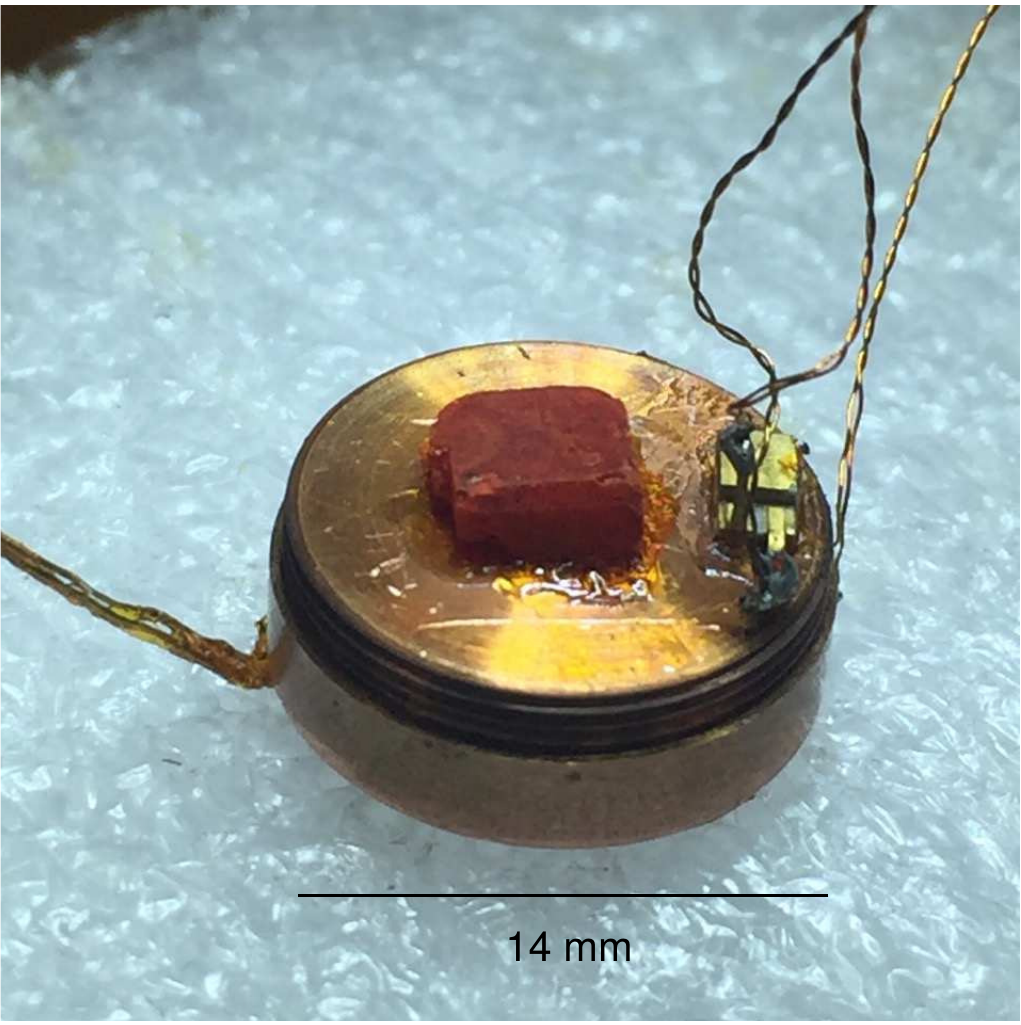}\caption{\label{fig:S_samples} TIPS-tetracene crystallites. Photograph of
Sample 3 mounted on sample holder.}
\end{figure}

\section{Samples}

Samples were prepared by evaporation from saturated solution. In the
main text, three samples were used which we refer to as samples 1-3.
These samples were crystallites of $\sim\text{mm}$ linear dimensions,
containing multiple domains (each of 100~$\mu\text{m}$ scale) with
a powder coating. Fig.~\ref{fig:S_samples} shows an image of Sample
3, Samples 1 and 2 were prepared in an identical way. X-Ray diffraction
confirmed that all samples indexed to the same unit cell previously
determined for TIPS-tetracene, demonstrating that they
had the same underlying solid-state structure.

In addition, a separate single crystal sample, which showed good extinction
under crossed-polarizers, was used to investigate the origin of the
broadening of the photluminescence resonances. These data are described
in Section \ref{sec:xtal} below.

\section{PL and $\Delta\mbox{PL}_{\mathrm{res}}$ spectra in samples 1-3}

\begin{figure}[h!]
\centering{}\includegraphics[width=1\textwidth]{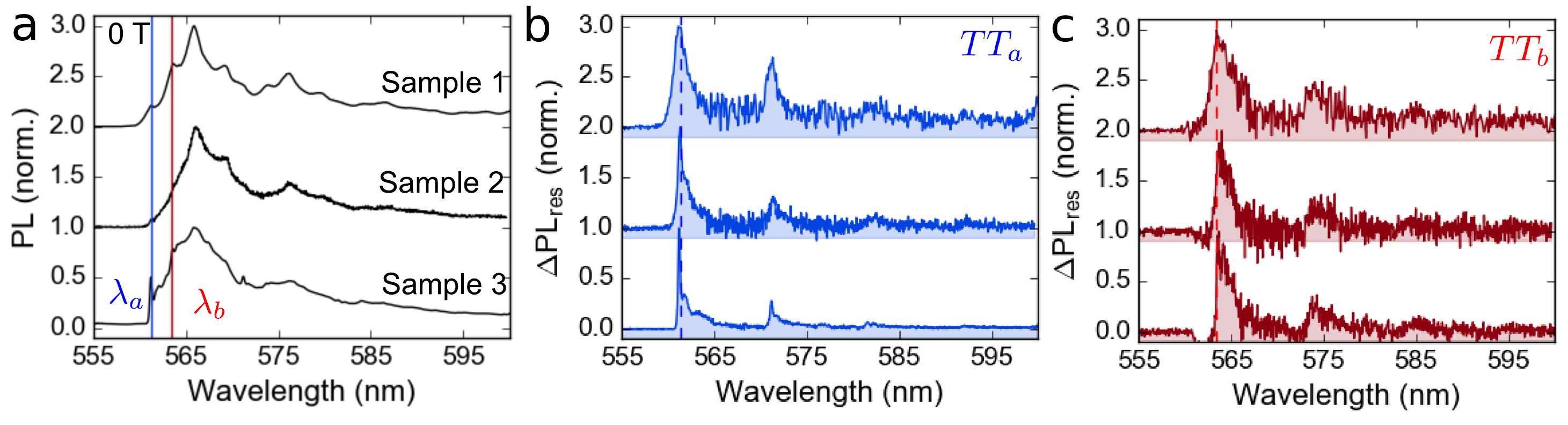}\caption{\label{fig:S1}Photoluminescence and triplet-pair site emission for
samples 1-3. \textbf{a} Zero-field emission with spectral positions
$\lambda_{\mathrm{a,b}}$ highlighted. \textbf{b} Extracted $TT_{\mathrm{a}}$
spectra using the 5.6 T resonance for samples 1-3 (top to bottom).
\textbf{c} Extracted $TT_{\mathrm{b}}$ spectra taken using the 2.98
T resonance for samples 1-3 (top to bottom).}
\end{figure}

Three samples, labelled 1-3, were used, as described in the main text.
Each sample was measured under slightly different conditions: Sample
1 and 3 were measured at 1.4 K and 1.1 K respectively, both immersed
in liquid helium, while Sample 2 was measured at 2~K under exchange
gas. Fig. \ref{fig:S1} shows that the relative intensities of $TT_{\mathrm{a-c}}$
in the bare PL vary between samples, as well as their extracted linewidths,
but not the peak positions/vibronic structure. We note that the grating
used for Sample 2 and 3 afforded higher spectral resolution than that
used for Sample 1, which contributes to differences in the observed
linewidths.

\section{Extraction of $\Delta\mbox{PL}_{\mathrm{res}}$ }

As described in the main text, resonant spectra were extracted by
taking the difference in PL on and off resonance: 
\begin{equation}
\Delta\mbox{PL}_{\mathrm{res}}=|\mbox{PL}(B_{\mathrm{res}.})-\mbox{PL}(B_{\mathrm{off\,res.}})|.
\end{equation}
Here we describe this procedure in detail for both cw and pulsed magnetic
fields.

\subsection{Resonant Spectra from Continuous Magnetic Field Sweeps}

For Sample 2 and 3, the magnetic field was continuously swept and
spectra continuously recorded. For a clean off-resonant subtraction,
$\mbox{PL}(B_{\mathrm{off\,res.}})$ was taken as the average of spectra
at symmetric points either side of each resonance: 

\begin{equation}
\Delta\mbox{PL}_{\mathrm{res}}=|\mbox{PL}(B_{\mathrm{res}.})-\mbox{PL}(B_{\mathrm{off\,res.}})|=|\mbox{PL}(B_{0})-\frac{1}{2}[\mbox{PL}(B_{0}+\Delta)+\mbox{PL}(B_{0}-\Delta)]|,
\end{equation}
where $B_{0}$ is the resonant field, and $\Delta$ the field separation
between the on-resonant and off-resonant points. To improve the signal-to-noise
ratio, the on-resonant and off-resonant spectra were averaged within
a smaller window $\delta$.

\begin{figure}[h!]
\centering{}\includegraphics[width=0.6\textwidth]{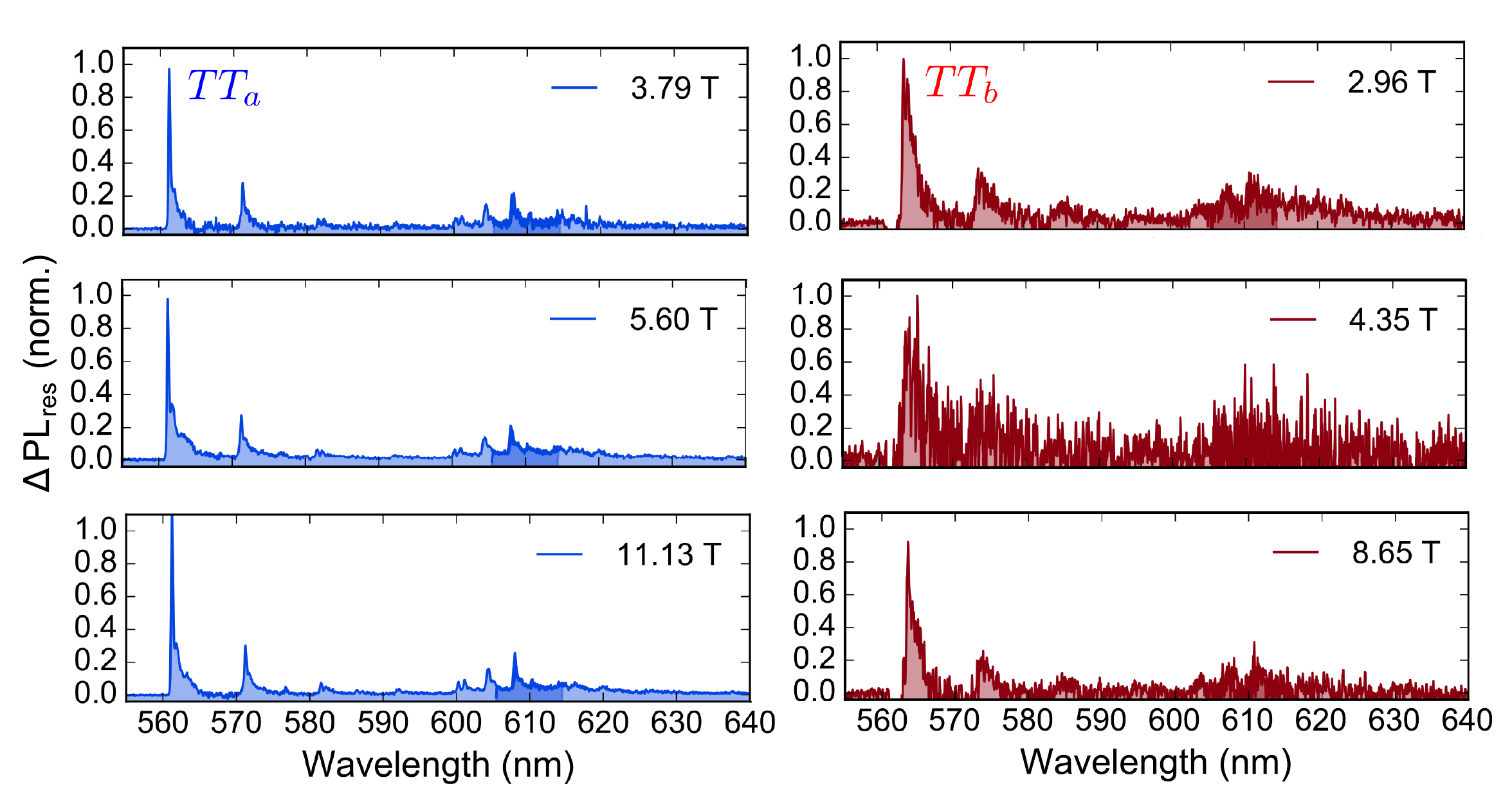}\caption{\label{fig:sample3}$\Delta\mbox{PL}_{\mathrm{res}}$ from continuous
field measurements taken on Sample 3 at 1.1 K. Parameters used to
extract spectra are detailed in Table 1.}
\end{figure}

The spectra associated with $TT_{\mathrm{a,b}}$ for each resonant
field are shown in Fig.~\ref{fig:sample3}. These show that the
same spectra can be extracted for each field in a given resonance
progression (3.79, 5.60, and 11.13~T for $TT_{\mathrm{a}}$ and 2.96,
4.35, and 8.65~T for $TT_{\mathrm{b}}$). The parameters used to
extract each spectrum are listed in Table~\ref{tab:sample3_params}.
Note that resonances that overlapped with neighbouring resonances,
such as the $4.25$~T and $8.63$~T dips, required smaller windows
for off-resonant spectra (smaller $\Delta$) and smaller averaging
windows ($\delta$).

\begin{table}[h!]
\centering{}%
\begin{tabular}{|c|c|c|}
\hline 
\multicolumn{3}{|c|}{Sample 3}\tabularnewline
\hline 
$B_{0}$ (T) & $\Delta$ (T) & $\delta$ (T)\tabularnewline
\hline 
\hline 
3.79 & 0.5 & 0.13\tabularnewline
\hline 
5.60 & 1.0 & 0.13\tabularnewline
\hline 
11.13 & 2.0 & 0.15\tabularnewline
\hline 
2.96 & 0.6 & 0.13\tabularnewline
\hline 
4.35 & 0.25 & 0.09\tabularnewline
\hline 
8.63 & 0.9 & 0.15\tabularnewline
\hline 
\end{tabular}\caption{\label{tab:sample3_params}Field positions and parameters used for
$\Delta\mbox{PL}_{\mathrm{res}}$ extraction for Sample 3. $B_{0}$
is the resonant field, $\Delta$ is the field step on either side
of $B_{0}$ at which non-resonant spectra are taken, and $\delta$
is the averaging window used for on and off-resonant spectral slices. }
\end{table}

\subsection{Resonant Spectra from Pulsed Magnetic Field Points}

\begin{figure}[h!]
\centering{}\includegraphics[width=1\columnwidth]{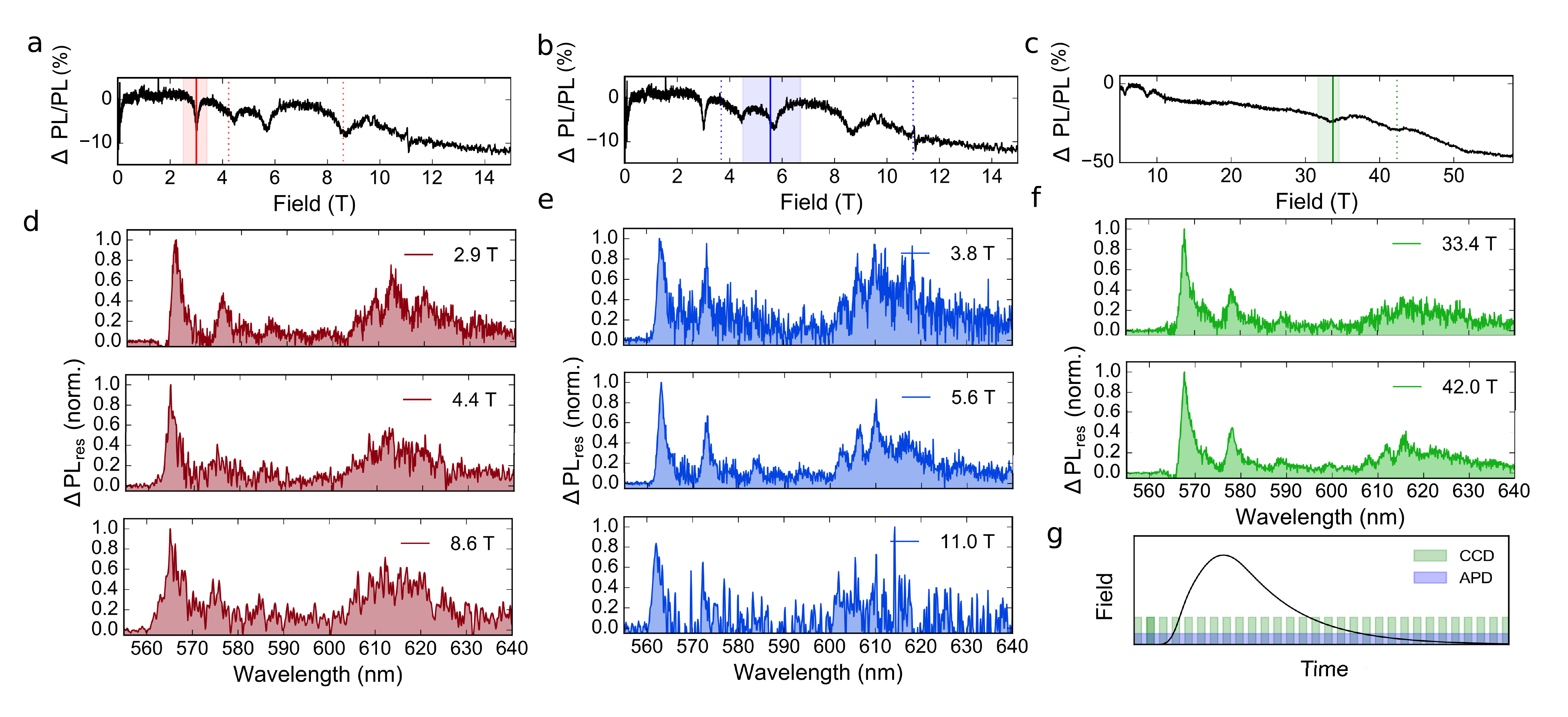}\caption{\label{fig:resspec}$\Delta\mbox{PL}_{\mathrm{res}}$ from pulsed
field measurements taken on Sample 1 at 1.4~K. \textbf{a} Field points
used for $TT_{\mathrm{b}}$ resonances: field point used in main text
(2.9~T, solid line) and remaining resonances at 4.4~T and 8.6~T
(dashed lines). \textbf{b} Field points used for $TT_{a}$ resonances:
field point used in main text (5.6~T, solid line) and remaining resonances
at 3.8~T and 11.0~T (dashed lines). \textbf{c} Field points used
for $TT_{c}$ spectrum: field point used in main text (33.7~T, solid
line) and remaining resonance at 42~T (dashed line). We note that
while the 33 and 43~T resonances have very similar spectral features,
they have distinct temperature dependences and so are associated with
distinct species. \textbf{d} $\Delta\mbox{PL}_{\mathrm{res}}$ for
all three $TT_{\mathrm{a}}$ resonances: 2.9, 4.4, and 8.6~T. \textbf{e}
$\Delta\mbox{PL}_{\mathrm{res}}$ for all three $TT_{\mathrm{\mathrm{b}}}$
resonances: 3.8, 5.6, and 11.0~T. \textbf{f} $\Delta\mbox{PL}_{\mathrm{res}}$
for the high-field resonances at 33.4 T and 42.0 T. \textbf{g} Schematic
of spectrally resolved PL measurements through pulsed field shots.
The black line represent the field through time, the blue gate shows
the continuous integrated measurement of the PL recorded on an avalanche
photodiode (APD) and the green gates show the shutter (2~ms) for
CCD exposure for spectrally resolved data. A series of shots to different
field points were used to extract $\Delta\mbox{PL}_{\mathrm{res}}$.}
\end{figure}

For measurements in pulsed magnetic fields (Sample 1), fields were
swept over $\sim100\,\mbox{ms}$ and so spectra could only be sampled
at discrete points (Fig~\ref{fig:resspec}g). In this case, symmetric
points about the resonances were not available for off-resonant subtraction.
To extract $\Delta\mbox{PL}_{\mathrm{\mathrm{\mathrm{res}}}}$, the
off-resonance spectra were therefore calculated via linear interpolation
between the available field points ($B_{\pm})$ either side of the
resonance at $B_{0}$:

\begin{equation}
\Delta\mbox{PL}_{\mathrm{res}}=|\mbox{PL}(B_{\mathrm{res}.})-\mbox{PL}(B_{\mathrm{off\,res.}})|=|\mbox{PL}(B_{0})-\frac{\Delta_{+}\mbox{PL}(B_{+})+\Delta_{-}\mbox{PL}(B_{-})}{(\Delta_{+}+\Delta_{-})}|
\end{equation}
\begin{equation}
\Delta_{+}=B_{+}-B_{0}
\end{equation}
\begin{equation}
\Delta_{-}=B_{0}-B_{-}
\end{equation}
Parameters used are in Table~\ref{tab:pulsed_params}. We found
good agreement with spectra extracted from both cw and pulsed field
measurements.

\begin{table}[h!]
\centering{}%
\begin{tabular}{|c|c|c|c|c|}
\hline 
Resonance (T) & $B_{0}$ (T) & $B_{+}$(T) & $B_{-}$(T) & $\Delta_{+}$, $\Delta_{-}$(T)\tabularnewline
\hline 
\hline 
2.98 & 3.049  & 3.362 & 2.488  & 0.313, 0.561\tabularnewline
\hline 
3.79 & 3.673  & 4.235  & 2.488  & 0.562, 1.185\tabularnewline
\hline 
5.60 & 5.501  & 6.753  & 4.466  & 1.252 1.035\tabularnewline
\hline 
4.36 & 4.23  & 6.406  & 2.22  & 2.176 2.01\tabularnewline
\hline 
8.63 & 8.614  & 9.398  & 7.82  & 0.784, 0.794\tabularnewline
\hline 
11.13 & 11.67  & 14.66  & 9.292 & 2.99, 2.378\tabularnewline
\hline 
33.4 & 33.78  & 34.55  & 31.7 & 0.77, 2.08\tabularnewline
\hline 
42.0 & 42.33  & - & 39.77 & 2.56\tabularnewline
\hline 
\end{tabular}\caption{\label{tab:pulsed_params}Field positions used for $\Delta\mbox{PL}_{\mathrm{res}}$
extraction for Sample 1. Spectra in Fig.~3 of the main text used
the resonances at 5.6, 2.98 and 32.4 T. Since it was not possible
to extract a spectral slice at $B_{+}$ for the 43.0~T resonance,
only a spectrum at $B_{-}$ was used for off-resonant subtraction.}
\end{table}

\section{Numerical Spectral Decomposition}

\begin{figure}[h!]
\centering{}\includegraphics[width=.7\columnwidth]{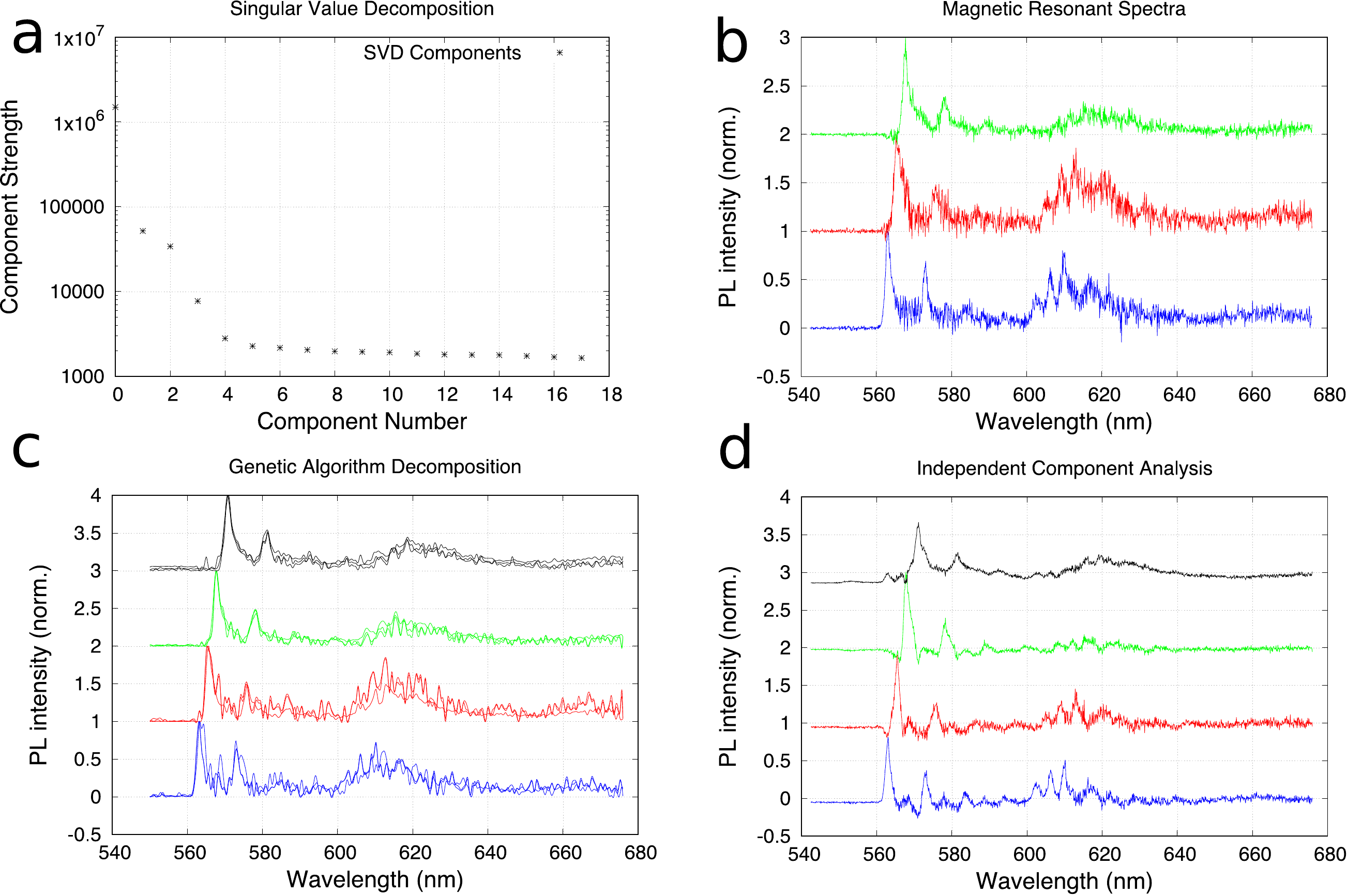}\caption{\label{fig:decomp}Spectral decomposition of magneto-photoluminescence
in Sample 2. \textbf{a} Singular value decomposition (SVD) of the
field-dependent spectra for pulsed field data up to 68 T, showing
four significant components. \textbf{b} Three species corresponding
to distinct exchange couplings (and hence level crossings) decomposed
using the magnetic level-crossing technique outlined above. \textbf{c}
Spectral deconvolution into four distinct components using a genetic
algorithm. \textbf{d} Spectral decomposition into four distinct components
using an independent component analysis algorithm.}
\end{figure}

To complement the approach used in the main text, and provide an alternative
means to extract triplet-pair optical signatures, the magnetic-field
dependent photoluminescence spectrum was decomposed into separate
components with machine-learning algorithms (Fig.~\ref{fig:decomp}).
We use two distinct algorithms which both reproduce the species observed
in the magnetic level-crossing technique. Using singular value decomposition,
four main components are present in the spectrum above the noise (Fig. S\ref{fig:decomp}a).
The spectra associated with those four components is then confirmed
using a genetic algorithm (documented previously \cite{Sgelinas2011binding}),
shown in Fig. \ref{fig:decomp}c, and separately using blind source
separation using an implementation of independent component analysis
in the python module ``Scikit-learn'' (Fig.~\ref{fig:decomp}c)
\cite{Sscikit-learn}. The results confirm the spectral separation
into three distinct magnetically active components (corresponding
to $TT_{\mathrm{a-c}}$), with the remainder of the spectrum showing
no distinct dependence on magnetic field.

\section{\label{sec:xtal}Magneto-PL of TIPS-Tetracene single crystal}

\begin{figure}[h!]
\centering{}\includegraphics[width=0.75\textwidth]{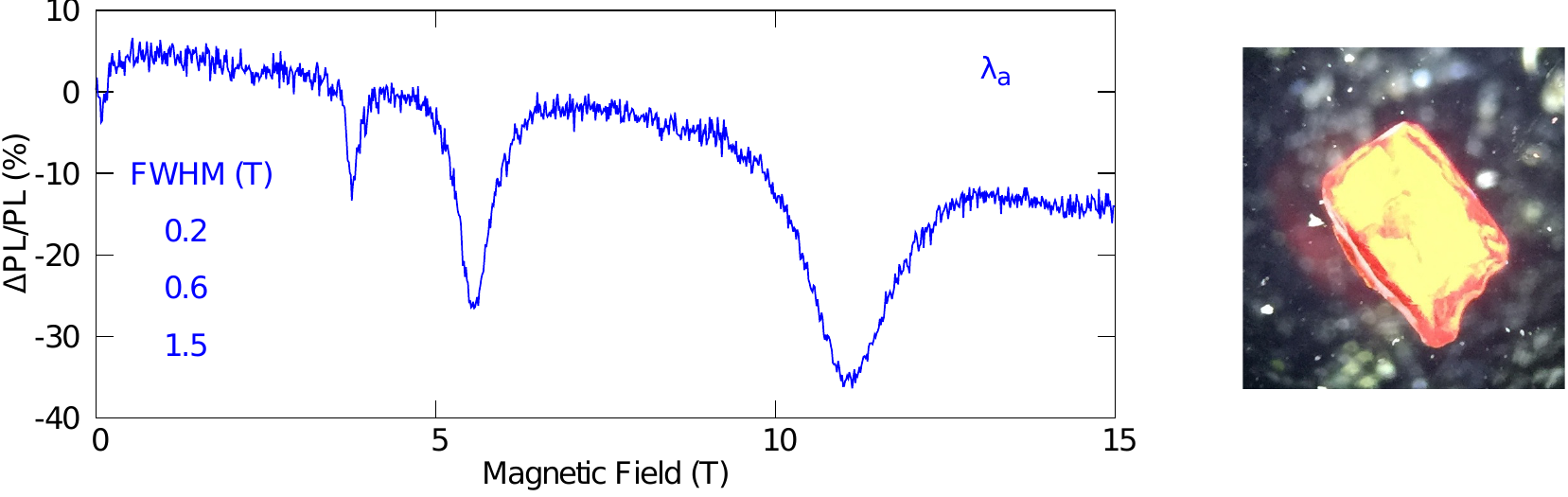}\caption{\label{fig:Sxtal}Left: $\Delta\mbox{PL/PL}$ for a TIPS-tetracene
single crystal at 1.1 K at the $\lambda_{\mathrm{a}}$ spectral position.
Excitation was at 488~nm and the orientation of the crystal axes
relative to the magnetic field was not determined. Right: crossed
polarizer image of single-crystal sample highlighting its optical
uniformity across the sample.}
\end{figure}
Fig.~\ref{fig:Sxtal} shows $\Delta\mbox{PL/PL}$ at $\lambda_{\mathrm{a}}$
in a single crystal TIPS-tetracene sample (confirmed using cross-polarized
optical microscopy - Fig. \ref{fig:Sxtal}, right), displaying comparable
magnetic linewidths to the multi-domain samples 1-3. This indicates
that the linewidths of the PL resonances are kinetically broadened,
instead of inhomogeneously broadened due to varations in exchange
coupling. This kinetic broadening is described in more detail below.

\section{Spin Hamiltonian and sign of $J$}

We adopt a triplet-pair spin Hamiltonian consisting of exchange, Zeeman
and zero-field splitting terms: 
\[
H=J\mathbf{S}_{1}\cdot\mathbf{S}_{2}+\sum_{i=1,2}(\mu_{B}\mathbf{B}\cdot\mathbf{g}_{i}\cdot\mathbf{S}_{i}+\mathbf{S}_{i}\cdot\mathbf{D}_{i}\cdot\mathbf{S}_{i})
\]
 where $\mathbf{g}_{1,2}$ and $\mathbf{D}_{1,2}$ are the $g-$ and
zero-field-splitting matrices for triplets 1 and 2. For the two triplets,
we assume that $\mathbf{g}$ and $\mathbf{D}$ differ only in their
orientations and not their principal values.\emph{ }Resonances occur
between the spin $S=0$ triplet-pair state $|S\rangle$ and one other
($S=2$ or $S=1$) state $|X\rangle$, forming an effective two-level
system with eigenstates $|1\rangle,|2\rangle$. The emission probability
for each state depends on their singlet projections $|\langle1|S\rangle|^{2}=\frac{1}{2}[1+(4V^{2}/\Delta^{2}+1)^{-1/2}]$,
$|\langle2|S\rangle|^{2}=1-|\langle1|S\rangle|^{2}$, where $V=|\langle X|H|S\rangle|$
is the mixing matrix element, $\Delta=\langle S|H|S\rangle-\langle X|H|X\rangle\simeq\frac{J}{2}S(S+1)-mg\mu_{B}B$
is the detuning with $m$ the spin projection along the field. This
gives resonant fields of $(1,\,1.5,\,3)|J|/g\mu_{B}$, independent
of the sign of $J$, and so here we choose $J>0$. Corrections to
the resonant field positions of order $D$ \cite{Sbayliss2016spin},
the triplet zero-field splitting parameter, could in principle be
used to extract the sign of $J$ with the sign of $D\,(>0)$ known
for TIPS-tetracene \cite{Sweiss2016strongly}. The linewidths of the
PL resonances currently make it challenging to extract any deviations
from the expected 1:1.5:3 field ratios, and so we leave this sign
determination for future work.

\section{Spin mixings}

Taking colinear $\mathbf{g}_{1,2}$, we find $|\langle T_{i}|H|S\rangle|=0$
making singlet-triplet mixings nominally forbidden for rotationally
equivalent triplets. For orientationally inequivalent triplets, high-field
spin-mixing becomes allowed with $g$-anisotropy through $|S\rangle-|T_{0}\rangle$
mixing. Taking $\mathbf{B}\parallel z$ we find $|\langle T_{0}|H|S\rangle/\Delta|=\sqrt{\frac{2}{3}}\mu_{B}B|\Delta g_{zz}|/J$,
setting a characteristic saturation field $\sim J/|\Delta g_{zz}|$
where $\Delta g_{ij}=g_{1,ij}-g_{2,ij}$. The singlet-triplet resonance
also becomes allowed through this mechanism, $|\langle T_{-}|H|S\rangle|=\frac{1}{\sqrt{3}}\mu_{B}B\sqrt{\Delta g_{zx}^{2}+\Delta g_{zy}^{2}}$.
Additionally, hyperfine interactions can mix singlet and triplet manifolds.
Since these have strength $\sim$mT , independent of the field, they
would not give rise to a high-field effect.

\section{Kinetic Modelling}

Simulations are as described in Ref. \cite{Sbayliss2016spin} which
uses the kinetic scheme outlined in Fig.~1\emph{A}, but incorporating
an anisotropic triplet $g$-matrix. Fig.~6\emph{B} uses the TIPS-tetracene
zero-field splitting parameter $D/g\mu_{B}=50\,\mbox{mT}$ \cite{SBayliss2014b,SBayliss2015b,Sweiss2016strongly},
a kinetic ratio (singlet reformation rate/dissociation rate) $\epsilon=10^{3}$
\cite{Sbayliss2016spin}, and a molecular $g$-matrix of $\mathbf{g}_{mol}=g\mbox{\ensuremath{\cdot}diag}(1-\delta,1,1+\delta)$
where $\delta=0$ or $2.5\times10^{-3}$. Simulations are for triplets
randomly oriented with respect to each other and the field. As outlined
above, mixing should only be effective for $V\sim\Delta$ giving linewidths
$\lesssim D$ for resonant fields $\lesssim$$50$ T. For large $\epsilon$
however, the resonant linewidths are increased as $\sim\sqrt{\epsilon}$,
providing an additional kinetic broadening. While the two emission
scenarios (from a separate singlet and from triplet pairs directly)
exhibit this kinetic broadening, in the case of triplet-pair emission,
$\Delta\mbox{PL/PL}$ reduces as $\epsilon^{-1}$ while for separate
singlet emission it saturates at 50~\%, indicating the presence of
triplet-pairs coupled to a site-specific emissive singlet. This is
described in more detail below.

\subsection{Singlet-Exciton Emission}

As described previously \cite{SMerrifield,Sbayliss2016spin}, we adopt
the following kinetic scheme for magneto-photoluminescence

\selectlanguage{american}%
\begin{equation}
\begin{array}{cc}
\begin{array}{ccc}
G & \searrow\\
 &  & S\\
 & \gamma_{r} & \downarrow
\end{array}\begin{array}{c}
\gamma_{+}\alpha_{i}\\
\rightleftharpoons\\
\gamma_{-}\alpha_{i}
\end{array}\begin{array}{c}
\Bigg\{\end{array}\begin{array}{c}
\\
\begin{array}{c}
P_{i}\end{array}\\
\\
\end{array}\begin{array}{c}
\Bigg\}\end{array}\begin{array}{c}
\gamma_{d}\\
\rightarrow\\
\\
\end{array} & T+T.\end{array}\label{eq:kinetics}
\end{equation}
Here $G$ is the generation rate of singlet excitons ($S$) and $\gamma_{r}$
their radiative decay rate. $\gamma_{\pm}$ are the rates of generation
and fusion of triplet-pair state $P_{i}$ and $\alpha_{i}=\langle P_{i}|P_{S}|P_{i}\rangle$
is their singlet content where $P_{S}=|S\rangle\langle S|$ and $|S\rangle$
is the spin-singlet state. When the effective fission rate is fast
compared to the radiative decay rate (i.e. fission is efficient),
the steady-state photoluminescence is
\begin{equation}
\mbox{PL}(B)\simeq\frac{G\gamma_{r}}{\gamma_{+}}\Big(\sum_{i}\frac{\alpha_{i}(B)}{1+\epsilon\alpha_{i}(B)}\Big)^{-1}
\end{equation}
\foreignlanguage{english}{where $\epsilon=\gamma_{-}/\gamma_{d}$.
For exhange-coupled triplets at zero-field, $\alpha_{i}=1$ for the
singlet state and is zero for all the other states. At the level crossings,
effective two-level systems are formed between }\foreignlanguage{british}{the
singlet state $|S\rangle$ and one other ($S=2$ or $S=1$) state
$|X\rangle$. This gives}\foreignlanguage{english}{ $\{\alpha_{i}\}=\alpha_{1},1-\alpha_{1}$
with }\foreignlanguage{british}{
\begin{equation}
\alpha_{1}=\frac{1}{2}\Big(1+\frac{1}{\sqrt{1+4V^{2}/\Delta^{2}}}\Big),
\end{equation}
where $V=|\langle X|H|S\rangle|$ is the mixing matrix element from
the spin-Hamiltonian $H$ and, $\Delta=\langle S|H|S\rangle-\langle X|H|X\rangle\simeq\frac{J}{2}S(S+1)-mg\mu_{B}B$
is the detuning with $m$ the spin projection along the field. This
yields
\begin{equation}
\frac{\Delta\mbox{PL}}{\mbox{PL}}=-\frac{\epsilon}{2(\epsilon+1)}\frac{1}{1+\Big(\frac{\Delta}{V\sqrt{2(\epsilon+2)}}\Big)^{2}}.\label{eq:dPLPL_s}
\end{equation}
}\foreignlanguage{english}{At resonance, $\Delta=0$ giving
\begin{eqnarray}
\frac{\Delta\mbox{PL}}{\mbox{PL}} & =- & \frac{\epsilon}{2(1+\epsilon)}\leq-\frac{1}{2},
\end{eqnarray}
setting a maximum $\Delta\mbox{PL}/\mbox{PL}$ of 50~\%, in the limit
$\epsilon\gg1$. From Eq. \ref{eq:dPLPL_s}, the resonant linewidth
(FWHM) is given by
\begin{equation}
\delta B=\frac{2V}{mg\mu_{B}}\sqrt{2(\epsilon+2)},
\end{equation}
which grows as $\sqrt{\epsilon}$ for $\epsilon\gg1$.}
\selectlanguage{english}%

\subsection{Triplet-Pair Emission}

If instead we consider triplet pairs emitting directly, the kinetic
scheme is
\begin{equation}
\begin{array}{cc}
\begin{array}{ccc}
G & \searrow\\
 &  & S\\
 & \gamma_{r} & \downarrow
\end{array}\begin{array}{c}
\gamma_{+}\alpha_{i}\\
\rightarrow\\
\\
\end{array}\begin{array}{c}
\Bigg\{\end{array}\begin{array}{c}
\\
\begin{array}{c}
P_{i}\end{array}\\
\alpha_{i}\gamma_{-}\downarrow
\end{array}\begin{array}{c}
\Bigg\}\end{array}\begin{array}{c}
\gamma_{d}\\
\rightarrow\\
\\
\end{array} & T+T,\end{array}
\end{equation}
and $\mbox{PL}=\gamma_{-}\sum_{i}\alpha_{i}P_{i}$. Repeating the
above procedure gives
\begin{equation}
\mbox{PL}=G\epsilon\sum_{i}\frac{\alpha_{i}^{2}}{(1+\epsilon\alpha_{i})}
\end{equation}
and
\[
\frac{\Delta\mbox{PL}}{\mbox{PL}}=-\frac{1}{(\epsilon+2)}\frac{1}{1+\Big(\frac{\Delta}{V}\sqrt{\frac{(\epsilon+1)}{(\epsilon+2)^{2}}}\Big)}.
\]
At resonance,
\begin{eqnarray}
\frac{\Delta\mbox{PL}}{\mbox{PL}} & =- & \frac{1}{\epsilon+2}\leq-\frac{1}{2}
\end{eqnarray}
which sets a maximum $\Delta\mbox{PL}/\mbox{PL}$ of 50~\%, but now
in the limit $\epsilon\ll1$. The resonance linewidths are
\begin{equation}
\delta B=\frac{2V}{mg\mu_{B}}\sqrt{\frac{(\epsilon+2)^{2}}{(\epsilon+1)}},
\end{equation}
which also increase as $\sqrt{\epsilon}$ for $\epsilon\gg1$.

Comparing the two scenarios of singlet exciton emission and triplet-pair
emission, only the former is able to reproduce the experimental observations
of simultaneously having kinetically broadened linewidths and changes
in photoluminescence approaching 50~\%. Our results therefore indicate
a scenario in which emission arises from singlet excitons, providing
a way of distinguishing these competing kinetic models.

\end{document}